# Optics of sunscreen lotions: Preliminary results on scattering and absorption coefficients.

Gunnar A. Niklasson[1], Sara L. Niklasson[1], Celina Notfors[2], Junxin Wang[1], Maria Strømme[2] and Cecilia Århammar[2,3]

[1] Department of Materials Science and Engineering, Solid State Physics, The Ångström Laboratory, Uppsala University, P.O. Box 35, SE-75103 Uppsala, Sweden.

[2] Department of Materials Science and Engineering, Nano Technology and Functional Materials, The Ångström Laboratory, Uppsala University, P.O. Box 35, SE-751 03 Uppsala, Sweden

[3] Disruptive Materials AB, Uppsala Science Park, 751 83 Uppsala, Sweden

## Abstract

Sunscreen lotions are used to protect the skin from damage due to solar ultraviolet (UV) radiation. The active UV-blocking components can be organic molecules or inorganic particles, for example $TiO_2$. While both *in-vivo* and *in-vitro* methods exist for assessing the protective capacity of sunscreens, their basic optical properties have received little attention. In this paper we take the first steps to address this issue by presenting a method to obtain spectra of absorption and scattering coefficients of a sunscreen by analysis of experimental transmittance and reflectance spectra. We model the optical properties of the sunscreen by two-flux theory, develop approximate expressions for interface reflection coefficients and obtain the remaining optical parameters by fitting the two-flux theory to measurements. We analyze spectrophotometric reflectance and transmittance spectra of experimental narrow band UV blocking materials based on mesoporous magnesium carbonate – $TiO_2$ mixtures, as well as some commercial sunscreens. Our method is demonstrated *in vitro* using quartz substrates and is further used to model *in vivo* measurements on the upside of the hand of one of the authors. The obtained absorption and scattering coefficient spectra give information on the relative contributions of these processes to the UV-blocking effect as well as the scattering of visible light responsible for the whitening of the skin upon application of a sunscreen lotion. Our analysis method may be useful for computational optimization of novel sunscreen formulations as well as for the development of *in-vivo* test methods.

# 1. Introduction

The effect of solar ultraviolet (UV) radiation leads to acute and chronic skin damage in humans, causing for example erythema (sunburn), immunosuppression, photo-carcinogenesis and photo-aging [1,2]. Therefore, the development of effective protective sunscreen lotions, creams and gels has become a high priority commercially as well as in applied research. The benefits of sunscreens for preventing skin cancer have been amply demonstrated [1]. A large number of UV-blocking materials are used in sunscreens and these can be classified as organic molecules and inorganic metal oxide particles [2,3]. Different organic filters exhibit high absorption and consequently give high protection in different parts of the UV-spectrum [2,3]. The inorganic filters in common use are currently limited to titanium dioxide ($TiO_2$) and zinc oxide (ZnO) [2,3]. They normally exhibit a broader absorption spectrum than chemical UV filters, and are limited by the band gap of the material. The organic filters primarily absorb UV radiation but the inorganic ones, in addition to high absorption, can display significant light scattering. In order to achieve broad spectrum protection in the whole UVB (290-320 nm) and UVA (320- 400 nm) wavelength ranges, an elaborate mix of filter materials is deemed to be necessary. The active molecules or nanoparticles are embedded in an aqueous solution or an emulsion consisting of a number of supplementary ingredients, such as emollients, emulsifiers and photo-stabilizers [3]. Typically, the active UV filters constitute 10-20% of the formulation, while water is the main component (>60%) [3]. In addition to providing UV blocking, minimization of visible light scattering is interesting for aesthetic reasons.

The sun protection factor (SPF) is a measure of the efficiency of sunscreens to protect against erythema. There exist standards for evaluation of the SPF of sunscreens based on *in-vivo* tests [4]. Basically, 2 mg/cm$^2$ of the sunscreen is applied to a patch of skin on a number of test subjects. This part of the skin is irradiated with UV-light from a solar simulator and subsequently the reddening of the skin is assessed, and a Minimal Erythema Dose response is determined. *In vitro* spectrophotometric determinations of SPF have encountered problems with reproducibility and reliability [5]. Uncertain factors include the difficulty of finding a substrate that mimics the interaction of the sunscreen with the skin as well as the repeatability of application and spreading of the sunscreen, and the need to establish and model the degradation of the SPF under UV irradiation [5]. However, an *in vitro* test procedure for UVA protection factor (UVA-PF) [6] has been introduced by the European cosmetics trade association COLIPA for evaluation of protection against UVA irradiation and a procedure for *in vitro* SPF measurement is currently under development [7]. This standard test uses roughened PMMA plates where the surface roughness is tailored to simulate the topographical features of the skin. A quantity comprising 1.3 mg/cm$^2$ of the sunscreen product is applied to the rough side of the PMMA plate and the UV transmission spectrum is measured. The procedure is repeated after exposure of the sample to a specified dose of UV radiation. After the *in-vitro* test, the appropriate protection factor is computed [3] by weighting the transmission over the solar irradiation spectrum (AM1.5) [8] and the erythema action spectrum [9], or alternatively the Persistent Pigment Darkening (PPD) spectrum [10]. Although correlations between the protection factors from *in-vivo* and *in-vitro* tests exist, especially for the UVA case, they are far from perfect [11,12].

In this paper we study the sunscreen formulation as a functional optical material and our objective is to determine optical parameters that can be used as input for calculations of the complex sunscreen-skin system. Current *in-vitro* test methods do not allow a material optics characterization of sunscreen lotions due to the complications induced by the rough substrates. An accurate method for determining absorption and scattering parameters of sunscreens would be desirable to put development of UV-blocking materials and formulations on a more quantitative basis. In addition, it is also interesting to quantify the visual appearance of the sunscreen on the skin, which can be studied by measuring light scattering in the visible range. For these purposes, a sample cell with well-characterized thickness and flat interfaces is needed. Surface roughness must be avoided in order not to introduce additional complications into the modelling of the optical response. We note that such an experimental configuration has been used as the "thin film method" of the Australian/New Zealand sunscreen product evaluation standard [13]. Here we use a similar method by enclosing the sunscreen between two UV-transparent quartz plates separated by a spacer. It should be emphasized that values of SPF obtained by the quartz cell method are not in general comparable to those obtained by standard test methods. This is because the surfaces of the quartz plates are flat and not rough as the skin; in addition, the evaporation of the water component is largely hindered by the sample cell and the degradation of the sunscreen under strong UV-irradiation is not considered. The effect of the latter two factors on sunscreen efficiency is well-known [14-16]. Our objective is to obtain absorption and scattering coefficients of the sunscreens, which can later be used in detailed optical models involving, for example, rough substrates and the features of the skin.

In this paper we present a method to obtain these coefficients from transmittance and reflectance spectra. The optical properties of the sunscreen layer are modeled by Kubelka-Munk (KM) two-flux theory [17,18]. Using experimental measurements as input data and fitting to the KM theory, it is possible to obtain values of the absorption and scattering coefficients of the sunscreen. Our method is based on a previous detailed study on the optical properties of nanoparticle suspensions, which showed that consistent inversion of the KM theory is possible [19]. However, additional approximations are necessary in the present case. First, we employ a diffuse scattering approximation, which means that regular transmittance is neglected, and specular reflectance is put equal to that of the front quartz interface. Secondly, we assume that the scattering pattern in the sunscreen lotion is peaked forward but isotropic in the backward direction. We put forward arguments showing that these approximations should be good in the present case.

In section 2 we briefly describe the sample preparation and optical characterization and motivate the diffuse scattering approximation. In section 3 we give a theoretical description of the KM formalism and the method used to invert these equations using experimental spectra as input. Section 4 describes experimental spectra and presents the obtained parameters for experimental UV-blockers based on mesoporous magnesium carbonate (MMC) – $TiO_2$ mixtures, as well as for some commercial sunscreens. The high surface area, narrow pore size distribution and proven hypoallergenic and bacteriostatic properties of MMC [20-23] makes it interesting as a component in sunscreen formulations. Finally, the results as well as possible future applications of the KM based computational method in the field of sunscreen materials research are discussed.

## 2. Sample preparation and characterization.

We first outline the preparation of the samples used in the optical measurements. Mesoporous magnesium carbonate (MMC) was synthesized by a procedure related to that described by Århammar *et al.* et al. [22] resulting in an opaque highly porous (Table S1) powder. The as-synthesized powder was hand-milled to particle sizes below 100 µm in order to improve spreadability of the formulation. Titanium oxide-MMC composites were synthesized by two different routes, without (synthesis 1) and with (synthesis 2) an acid catalyst (HCl, Sigma-Aldrich, 37% ACS reagent grade). Briefly the synthesis started from a solution of Titanium tetra-isopropoxide (Sigma-Aldrich, 97%) and 1-propanol (Sigma-Aldrich, analytical standard). Subsequently, MMC was added to this solution, which was then stirred and left to dry, resulting in a white (synthesis 1) or opaque (synthesis 2) powder. The powder obtained by Synthesis 1 was found to consist of MMC with amorphous $TiO_2$ in the pores (Table S1) and with additional spherical $TiO_2$ particles sized mainly in the µm region (Fig. S1, S2). The powder in synthesis 2 instead contained larger flakes of $TiO_2$ ranging from a few to hundreds of µm (Fig. S2). Prior to formulation, the $TiO_2$-MMC particles were hand-milled to improve sensory properties of the materials. A brief overview of the synthesis procedure and the characterization of these powders [22] is presented in Supplementary Information (SI). For comparison, crystalline $TiO_2$ (Degussa P25) nanoparticles (forming µm-sized agglomerates in the formulation) was also used as a sun-blocking pigment. Experimental UV-blocking emulsions were prepared by mixing 10 wt% of the MMC-$TiO_2$ syntheses 1 and 2 and Degussa into Propyless (20% propylene glycol; Apoteket AB, Sweden). We have also studied commercial sunscreens, acquired from local vendors. They are here denoted as Sunscreen A, B and C. Sunscreen A has the lowest SPF, as given by the manufacturer, while Sunscreen C exhibits the highest one. Sunscreen A only contains organic sun-blocking compounds, while B and C contain $TiO_2$ in addition to various organic compounds (Table S2).

Samples of the commercial sunscreens for optical measurements were prepared by placing approximately 0.05 g of the lotion onto a quartz plate with 11 µm thick aluminum foil spacers on both ends. Subsequently, another quartz slide was pressed onto the sample, allowing excess material to escape at the sides through the spacing between the two slides. This procedure results in samples with reproducible optical properties and a well-defined sample thickness of about 11 µm. The a-$TiO_2$/MMC samples for optical measurements were prepared by placing approximately 0.05 g of sample onto a quartz plate. On top of the sample another quartz plate was pressed in order to obtain a flat top surface and even thickness. These samples were highly viscous and no material escaped through the spacing between the two slides. Hence the sample thickness was estimated from the amount of material and sample area.

Total reflectance ($R$) and transmittance ($T$) in the wavelength range from 250 to 800 nm, using a step size of 1 nm, were measured by a Perkin Elmer Lambda 900 spectrophotometer, equipped with an integrating sphere. The optical experiments were carried out using a Spectralon reflectance standard. A correction procedure based on the calibrated reflectance of the standard, as well as a factor to account for port losses [24], was applied to all the data. For the sunscreen with the highest visible transmittance, diffuse reflectance and transmittance were measured by letting specular reflected light escape from the integrating sphere and absorbing the regular transmitted light, respectively. By comparison with the total $R$ and $T$, the regular transmittance and specular reflectance could be estimated. However, because of the geometry of the integrating sphere also some low angle scattered light is included in the measured regular $T$ and specular $R$ [19], so these results should be used with caution.

## 3. Theory

The optical properties of a light scattering material can be described in terms of the spectral absorption and scattering coefficients. These coefficients are material dependent and depend also on the size, shape and concentration of the scattering units together with the refractive index contrast between them and the embedding matrix. The inhomogeneities giving rise to scattering may be large molecules, or particles with sizes in the nano- and micrometer ranges. For particle sizes approaching the wavelength of light, light scattering becomes strong and must be taken into account to describe the optical properties. This is the regime of multiple scattering in a composite material, which is usually treated in the framework of radiative transfer and related theories [25,26]. The Kubelka-Munk (KM) two-flux model [17,18] is an analytical approximation to full radiative transfer, that is frequently used in applied research. In this paper we are concerned with the inverse problem of KM theory, i.e. the determination of the absorption ($K$) and backscattering ($S$) coefficients from spectrophotometric measurements of reflectance and transmittance. KM theory can be viewed as a starting point for ongoing efforts to solve the inverse problem of full radiative transfer [27,28] but even in the KM case the inversion procedure is quite complex [19].

The Kubelka-Munk model [17,18] considers two radiation fluxes moving in the forward and backward direction in a layer of material, as shown in Fig. 1. The coordinates $z=0$ and $z=d$ represent the front and back interfaces of the layer, and $d$ is the thickness of the sunscreen layer. The layer is assumed to be normally illuminated with light of intensity $I_0$. The total intensities in the forward and backward directions are coupled according to the differential equations:

$$dI/dz = -(S+K)I + SJ \qquad (1)$$

$$dJ/dz = (S+K)J - SI \quad . \qquad (2)$$

Here the intensity passing towards the back interface is denoted $I$ and the intensity towards the front interface is denoted $J$. The solutions of these equations give the transmittance and reflectance for the case of a slab with backside reflectance $R_g$ (Figure 1), according to,

$$R_{KM} = \frac{1-R_g[a-b\coth(bSd)]}{a-R_g+b\coth(bSd)} , \qquad (3)$$

and

$$T_{KM} = \frac{b(1-R_g)}{(a-R_g)\sinh(bSd)+b\cosh(bSd)} , \qquad (4)$$

where $a$ is equal to $(1+K/S)$ and $b$ is equal to $(a^2-1)^{0.5}$. After considering reflections at the front interface together with multiple reflections in the slab, we obtain

$$R = R_c + \frac{(1-R_c)(1-R_j)R_{KM}}{1-R_j R_{KM}} , \qquad (5)$$

and

$$T = \frac{(1-R_c)T_{KM}}{1-R_j R_{KM}} , \qquad (6)$$

where $R_c$ is the reflectance of the incident light on the front interface of the slab while $R_j$ is the front-side internal reflectance of the slab (Figure 1). It is realized, that in order to invert the equations for $R$ and $T$ to determine $S$ and $K$, knowledge of the interface reflectance parameters is necessary.

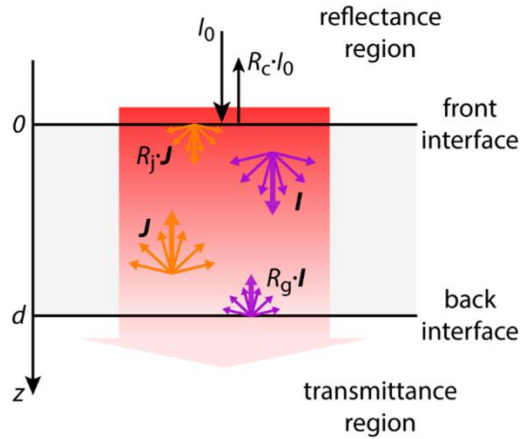

Figure 1. Schematic illustration of light scattering in a layer of thickness $d$, showing the parameters of the Kubelka-Munk two-flux model.

We now discuss the application of this formalism to the case of measurements on sunscreen lotions. In our measurements, the sunscreens are placed between two quartz glasses, which is a more complex sample configuration, than the one assumed above. The refractive index of quartz varies from 1.51 to 1.45 in the wavelength range 250-800 nm [29]. The sunscreen formulations contain mainly water and organic compounds and only smaller amounts of high refractive index inorganic compounds. The refractive index of water varies between 1.40 and 1.33 at UV and visible wavelengths [30] and organics usually have refractive indices in the 1.5 to 1.6 range. The refractive index of MMC has to this date not been determined, but anhydrous crystalline magnesium carbonate has a refractive index of 1.5 to 1.7, depending on crystal orientation [31]. The different polymorphs of $TiO_2$ have reported refractive index from 2.4 to 2.6 [32]. The refractive indices of the organics, water and MMC are sufficiently close to the values for quartz so that interface reflections between quartz and sunscreen lotion can be neglected to a good approximation. The refractive index of titanium dioxide is higher, but it is present as particles to a content of maximum 10 wt%.

In order to determine the interface reflectances further approximations are necessary. All our samples are whitish or visually opaque to the eye. This indicates that we can use the assumption that no light is propagating directly through the sunscreen layer. Instead, all the light is either scattered or absorbed in this layer. This was checked by supplementary measurements on the sample with the highest visible transmittance (Sunscreen A). It was found that the regular transmittance was less than 0.01 in the whole studied wavelength range, and this number includes also some low angle scattering. The specular reflectance was close to the value expected for the air-quartz interface. This means that the transmittance is completely diffuse to a very good approximation and the front side reflectance $R_c$ can be obtained from the refractive index of quartz [29].

Determination of the other interface reflectances requires in principle knowledge of the angular distribution of scattered light in the sunscreen layer [19,33]. In the absence of angle resolved scattering measurements certain simple approximations have been developed previously [19].

The simplest approximation would be to assume an isotropic scattering profile for both forward and backward directions. However, in this case the inversion procedure does not converge for our experimental data, i.e. it is not possible to fit exactly the measured $R$ and $T$ by KM theory for any combination of $S$ and $K$ coefficients. Instead, for the so-called **cri/dif** approximation [19], the inversion of KM theory converged for all our samples, giving exact fits to the experiments. In this approximation we assume that the scattering is anisotropic and exhibits a peak in the forward direction, while the backscattering is isotropic. This approximation can be rationalized if we assume that the scattering centers have sizes of the order of or larger than the wavelength. In this case dielectric particles exhibit scattering patterns that are strongly peaked in the forward direction, but more isotropic in the backward direction [34]. A strongly peaked distribution can be approximated by a constant intensity of scattered light out to the critical angle of total internal reflection and zero intensity at larger angles [19]. In this case we obtain the following expressions for the internal interface reflectances:

$$R_g = \frac{\int_{0°}^{\theta_c} r(\beta) \sin(2\beta) d\beta}{\int_{0°}^{\theta_c} \sin(2\beta) d\beta}, \qquad (7)$$

and

$$R_j = \frac{\int_{0°}^{90°} r(\beta) \sin(2\beta) d\beta}{\int_{0°}^{90°} \sin(2\beta) d\beta}. \qquad (8)$$

Here *r* denotes the reflection coefficient for light incident on the interface from inside the composite material at an angle $\beta$. It was computed from the refractive index of quartz, hence neglecting reflections between quartz and sunscreen, as argued above. The critical angle $\theta_c$ is the angle above which total internal reflection occurs so that the reflection coefficients for *s*- and *p*-polarized light are equal to unity. It was put equal to 42.5°, which is an average for quartz in the UV-visible range. The critical angle exhibits a slight wavelength dependence, which has a negligible influence on our analysis.

## 4. Results

In this section we present spectra of Kubelka-Munk coefficients of several experimental and commercial sunscreen formulations, obtained from inversion of experimental total transmittance and reflectance measurements. We first consider results for UV-blocking materials based on MMC/TiO$_2$. Figure 2 presents transmittance and reflectance of two a-TiO$_2$/MMC composites as well as Degussa TiO$_2$, all dispersed in Propyless.

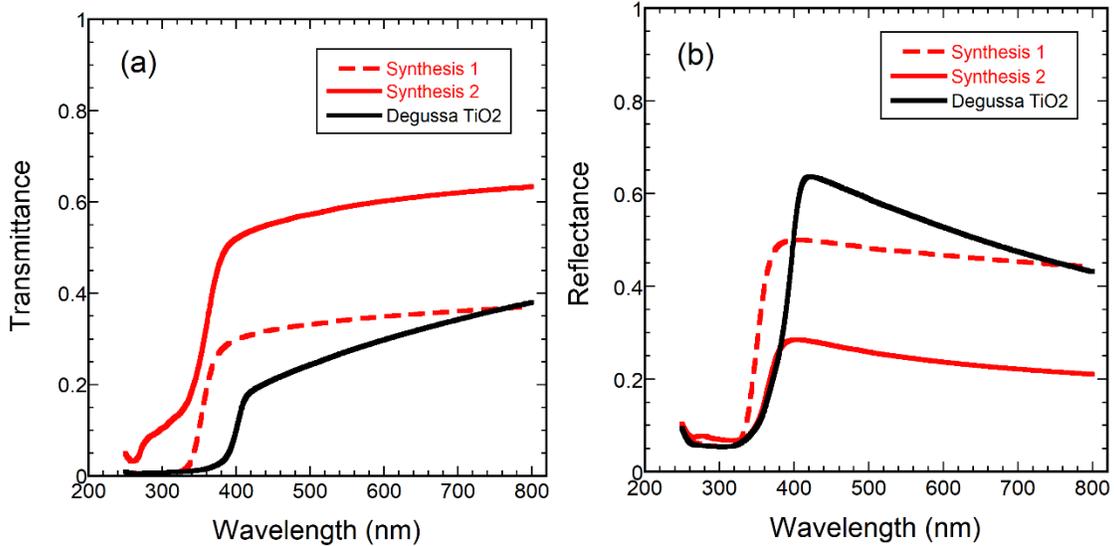

Figure 2. Total transmittance (a) and reflectance (b) measured on MMC-TiO$_2$ (Synthesis 1, 2) and Degussa TiO$_2$ dispersed in Propyless and enclosed between two quartz slides. Sample thicknesses were estimated to be 85 μm.

As discussed above, the transmittance is dominated by the diffuse component, because of the strong light scattering of the layers. The transmittance is high in the visible region, especially for Synthesis 2 and displays a sharp decrease in the ultraviolet due to the fundamental band to band absorption. It is seen that MMC-TiO$_2$, especially Synthesis 1, shows significant absorption of UVB radiation. Degussa TiO$_2$ is a good UVB-absorber and absorbs strongly also UVA radiation. Thus, the MMC-TiO$_2$ formulations may be used for blocking UVB, but should in practical applications be complemented by UVA blockers. The reflectance shows a high peak on the low energy side of the bandgap, which is most prominent in the case of the Degussa TiO$_2$ sample. The high diffuse reflectance extends into the visible region and causes the whitening effect observed when using sunscreens with high amounts of TiO$_2$.

Figure 3 presents absorption and backscattering coefficients obtained from inversion of the Kubelka-Munk theory, as discussed in the Theory section. The absorption coefficient exhibits a steep edge in the UV region due to the fundamental band gap of the materials. It is observed that the MMC-TiO$_2$ materials exhibit bandgaps in the range of 3-3.5 eV (wavelengths of 340-400 nm). The absorption coefficient in the visible range is low and it may, at least in part, be spurious due to scattered light escaping from the edges of the samples. The backscattering coefficient increases towards shorter wavelengths in the visible range, as expected for light scattering from particles, and then exhibits a sharp decrease correlated with the absorption edges. Hence the light scattering in the UV is diminished by the strong UV absorption in these materials. However, scattering is still significant in a large part of the UV and certainly influences the optical properties also in this range.

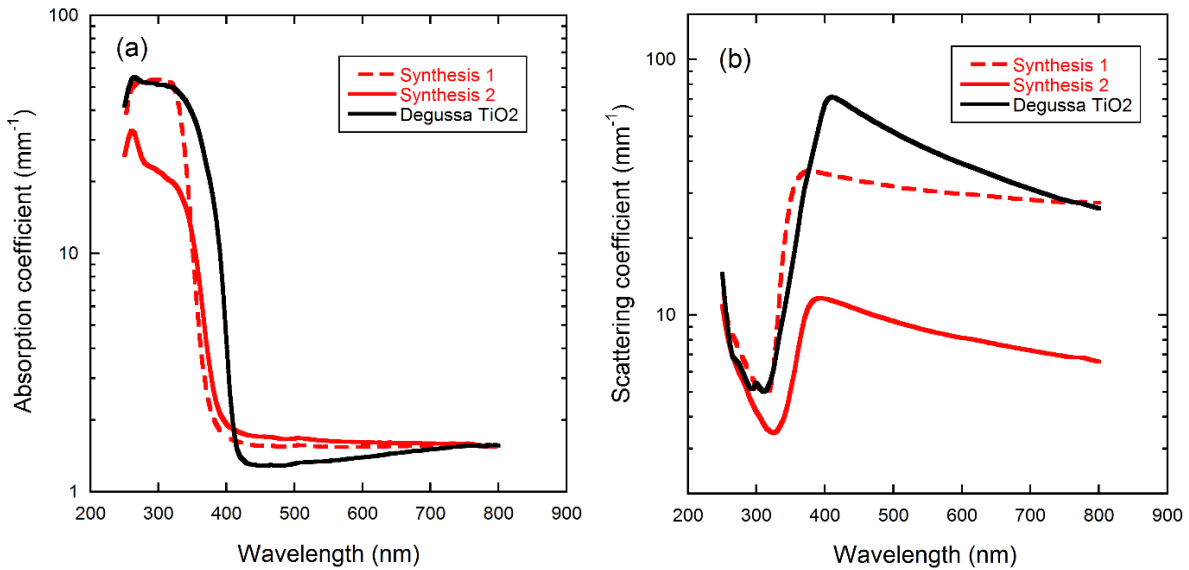

Figure 3. Absorption (a) and backscattering (b) coefficients as a function of wavelength for MMC-TiO$_2$ (Synthesis 1, 2) and Degussa TiO$_2$ dispersed in Propyless and enclosed between two quartz slides.

Figure 4 shows the experimental spectral transmittance and reflectance of three commercial sunscreen products. Sunscreen A is a low SPF product with a very high transmittance, approaching 0.9, in the visible range. The transmittance exhibits an abrupt decrease towards the UV, close to the onset of the UVA region. The structure in the spectrum will be discussed below. Sunscreen B also displays a rather abruptly decreasing transmittance at 400 nm. These spectra highlight the high accuracy of tailoring the onset of absorption to the onset of the UV region that is achieved by commercial sunscreen manufacturers. Sunscreen C is a high SPF product and its UV transmittance is very low, of the order of 0.003 to 0.01. The reflectance shows a similar pattern as in Figure 2, with a light scattering peak at a wavelength close to the transmittance decrease.

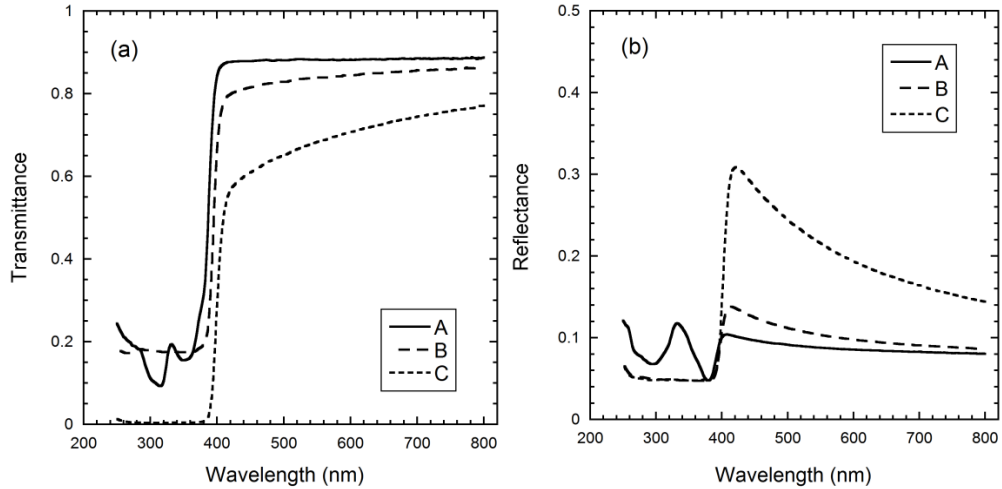

Figure 4. Total transmittance (a) and reflectance (b) of three commercial sunscreens enclosed between two quartz slides. The thicknesses of these sunscreen layers were about 11 μm.

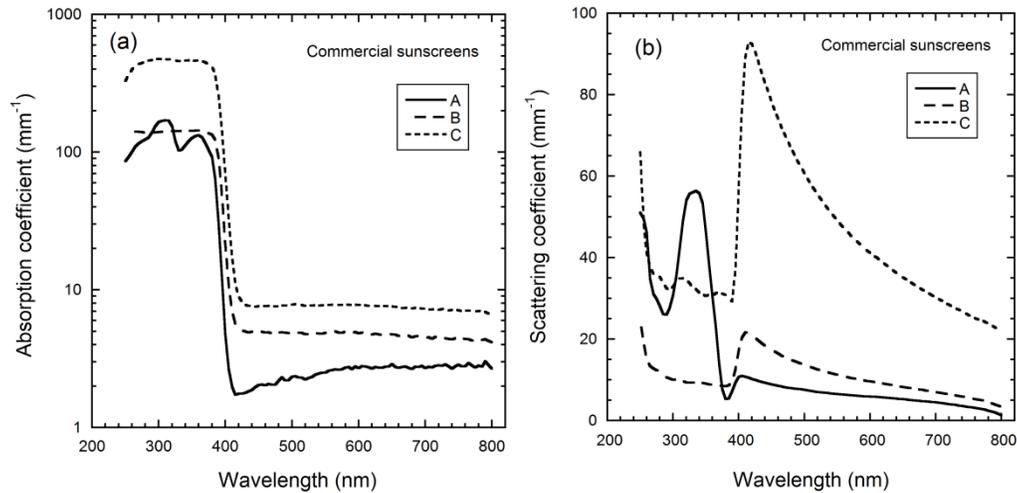

Figure 5. Absorption (a) and backscattering (b) coefficients as a function of wavelength for three commercial sunscreens labelled A, B and C, positioned between two quartz slides.

Figure 5 shows the absorption and backscattering coefficients obtained from fitting the spectra in figure 4. The absorption is very high throughout the UV region, but the backscattering is clearly significant, at least for Sunscreens A and C. Light scattering increases the path length of light traversing the sample and hence leads to a stronger absorption than predicted from the intrinsic properties of the absorbing material alone. This effect is partly responsible for the very high absorption coefficient of Sunscreen C. Hence the effect of scattering cannot be neglected when formulating optical models for sunscreens in the UV range. Sunscreen A exhibits two clear absorption peaks, approximately correlated to minima in the backscattering coefficient. These correspond to absorption peaks in the spectra of Ethylhexyl Triazone (~315 nm) and Butyl Methoxydibenzoylmethane (~360 nm), respectively [15]. The third UV filter used in Sunscreen A, Sodium Phenylbenzimidazole Sulfonate, provides further UVB absorption, exhibiting a main absorption peak at 302 nm [35]. In sunscreen B, $TiO_2$ contributes to the featureless absorption in the UVB and short wavelength UVA-region, the presence of $TiO_2$ is also signified by the sharp peak in the backscattering coefficient. Bis-Ethylhexyloxyphenol Methoxyphenyl Triazine and Butyl Methoxydibenzoylmethane contribute to the UVA-absorption, increasing the absorption coefficient in the long wavelength UVA-region. Sunscreen C contains a larger number of UV-blocking filters, including $TiO_2$, (Table S2) and exhibit a very uniform absorption throughout the UV region.

## 5. Discussion

In the previous sections we have described a method to determine the optical parameters of light scattering materials and applied it to the case of sunscreen lotions. Their optical properties are strongly dependent on the absorption and backscattering coefficients of the sunscreen formulations. In this section, we discuss the importance of this information for development of sunscreens as well as for gaining insight into new *in vivo* testing methods.

In the development of sunscreen products, a useful first step is to compute the efficiency of a formulation from known or independently measured absorbance spectra of the UV-blocking organic and inorganic filter materials [36,37]. This so called *in silico* calculation is usually based on irregular sunscreen film models, in order to take into account the statistical thickness distribution of the sunscreen when applied onto the rough surface of the skin [12,38]. We envisage that these models may be improved by using, as input to a KM-based model, independently measured absorption and scattering coefficients of solutions containing the constituents of a sunscreen lotion. Effects of scattering might be important also in the UV, at least in some cases, and thus theories taking this into account could be of considerable value.

The development of spectroscopic methods for *in vivo* testing of sunscreens is another topic of current concern. Currently used *in vivo* tests deliver UV irradiation to test subjects in doses large enough to cause erythema, which raises ethical concerns. In contrast, measuring the UV spectrum of the sunscreen-skin system has been estimated to deliver a dose level of only 1.6% of the erythemal relevant one [39]. A proposed new method calculates the UVA-PF from spectral measurement on the skin of test subjects with and without applied sunscreen. However, the optical modelling is very simplified and the "sunscreen transmittance" is estimated from the simple equation [40]

$$T(\lambda) = \sqrt{\frac{I(\lambda)}{I_0(\lambda)}} \quad , \tag{9}$$

where $I_0$ denotes the measured intensity from the skin and $I$ is the intensity in the presence of sunscreen. This equation neglects the surface reflection from the sunscreen layer and assumes that the skin reflectance is the same in both measurements. The latter assumption is dubious since refractive indices of sunscreen and skin are probably rather close to one another. It seems that the equation can be rationalized only if the interfacial reflectance values are low and the optical properties of the sunscreen layer are dominated by absorption. It appears to be less correct in cases (for example powders) where the functional property depends to a large extent on scattering. The method based on Eq. (9) has been used only for the UVA part of the spectrum, but it can be extended to testing of SPF by matching the UVA spectrum from the test subject to the spectral shape of an *in vitro* spectrophotometric measurement on the sunscreen [41]. Quite impressive correlations of the method to *in vivo* erythemal tests of UVA-PF and SPF have been demonstrated [39,41,42], provided that photostability is properly taken into account [41].

To increase the understanding of this test method, we performed a simple *in vivo* test measurement. Figure 6 shows the total reflectance of a portion of the upside of a hand before and after applying a specified amount of sunscreens A and B. The applied sunscreen layer was made sufficiently thick in order to clearly observe the visible whitening effect. First, we note that the reflectance with applied sunscreen is actually higher than the skin reflectance in the visible wavelength range (Figure 6a). The reason is that the refractive indices of sunscreen and skin are close and hence it is difficult to optically define an interface between the sunscreen and the skin. The similarity between the reflectance curves in the visible range indicates low interface reflection between these layers. Instead, we have two layers characterized by different $S$ and $K$ coefficients. The increase of the reflectance upon applying sunscreen is visible to the eye as a whitening of the skin and indicates that the backscattering coefficient of the sunscreen layer is higher than that of the underlying skin.

However, the situation in the UV is different, since the skin surface reflectance is low and absorption is dominating over scattering in this wavelength range. In principle we can use our determined absorption and backscattering coefficients in the KM reflectance expressions, Eqs. (3) and (5), putting $R_g$ equal to the reflectance of the underlying skin ($R_c$ and $R_j$ being determined as above), to model the UV part of the spectra in Figure 6. This constitutes a scientifically motivated improvement to Eq. (9), which accounts for the effects of light scattering. The results are given in Figure 6b, which shows a good correspondence between theory and measurements in the UV range. The reflectance of sunscreen B has already saturated to the surface reflectance of the sunscreen layer and therefore comparison with theory is not sensitive to input data in this case. However, the essential spectral features of sunscreen A, applied on the skin, are well reproduced by the KM calculation. This indicates that the KM model, although still very simplified, could be a starting point for formulating more detailed descriptions, involving skin topography as well as properly accounting for the skin-sunscreen interface. In part of the UV region, both the experiments and calculations for sunscreen A show an increase in reflectance above that of the skin. This effect points to a potential problem with the simple Eq. (9), also when its use is restricted to the UVA region.

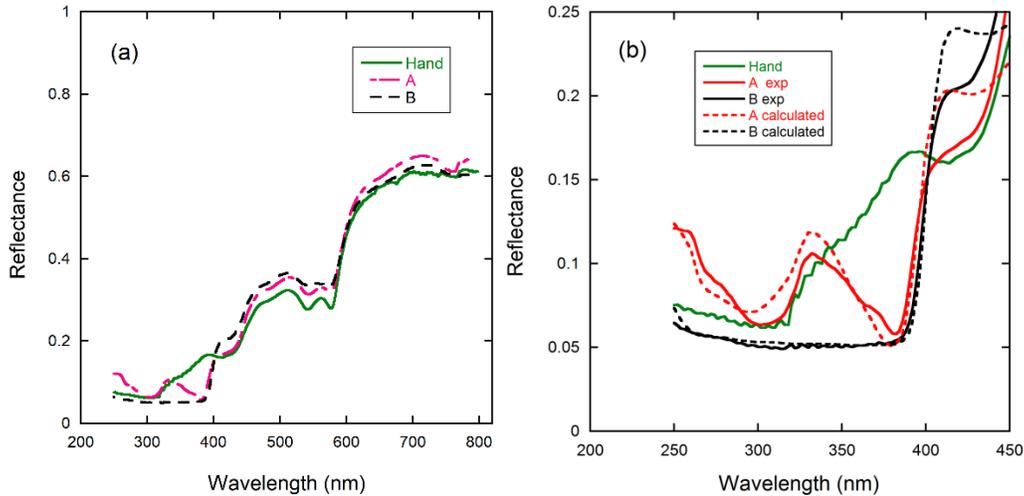

Figure 6. (a) Total reflectance of the upside of the hand of one of the authors (SLN) before and after application of an amount of 3 μL/cm$^2$ of sunscreens A and B. (b) Total reflectance in the UV-region compared to calculations by the Kubelka-Munk model, Eqs. (3) and (5). In the calculations a film thickness of 30 μm was assumed.

Given these potential problems, why does the spectrometric *in vivo* method work in practise? A precise answer requires improved optical modelling, but some experimental features of the method may be of importance. The method uses two optical fibre probes, positioned closely together, for illumination and detection, respectively [41,42]. The probes are separated from the skin by a protective Teflon spacer [41][41], which probably leads to partial refractive index matching between probe and sunscreen/skin, and hence to a decreased surface reflection. This effect could lead to an increased dynamic range in the ratio in eq. (9). Secondly, the surface reflection might not be detected by the probe and only the backscattered light close to the incident direction is detected. In this geometry, the absorption would dominate the response more than in an integrating sphere measurement, and effects of scattering would be less important. However, to verify the new *in vivo* test method, improved optical models pertinent to the experimental configuration are needed, including a parametrization of the optical properties of the skin as well as the sunscreen. We envisage that, to improve the understanding of the pros and cons of spectrometric *in vivo* testing, information of absorption and backscattering coefficients of the sunscreen layers will be very valuable. We emphasize that the development of such test methods is a pressing concern for ethical reasons.

## 6. Conclusion

In this paper we study sunscreen lotions as functional optical materials. We show how to determine absorption and backscattering coefficients of a sunscreen layer from spectrophotometric measurements by solving the inverse problem of Kubelka-Munk theory. These parameters determine the functional optical response of such layers, although sunscreen-skin interactions complicate matters in practical testing. We have shown the applicability of our method to experimental UV-blockers based on MMC with added titania pigments, as well as to commercial sunscreens. The a-$TiO_2$/MMC composite obtained from Synthesis 2 exhibits a lower UV absorption and a lower light scattering than the Degussa $TiO_2$ formulation. The a-$TiO_2$/MMC composite obtained from Synthesis 1 has a UV absorption and light scattering closer to that of $TiO_2$. Light scattering dominates the response in the visible wavelength range and leads to an increase in skin reflectance upon application of sunscreen. This visually observed skin whitening effect correlates with the measured reflectance and in addition the scattering coefficient in the visible region is signified by a peak on the low energy side of the bandgap. Absorption is dominating in the UV region, although scattering cannot be neglected and gives a significant contribution to the optical spectra in many cases. Detailed knowledge of the optical absorption and backscattering coefficients of the constituents of a sunscreen is of importance for the development of new sunscreen formulations. These optical parameters are also required for developing improved optical models of the sunscreen-skin system, which are needed to validate new *in vivo* test methods.

## Acknowledgments

We acknowledge the Swedish Energy Agency and the Swedish Science Council for funding.

Supplementary Information

**Preparation and characterization of MMC and MMC/TiO$_2$ composite samples.**

More detailed information can be found in ref. [22].

**Materials**

MMC was synthesized according to the procedure described by Cheung *et al.* [20], resulting in a transparent powder with a measured surface area of 443 m²g$^{-1}$ and a pore volume of 0.99 cm³g$^{-1}$. Titanium tetra-isopropoxid (TTIP) (Sigma-Aldrich, 97%), 1-propanol (Sigma-Aldrich, analytical standard), hydrogen chloric acid (HCl) (Sigma-Aldrich, 37% ACS reagent grade) and deionized water were used as sol-gel reagents. As purchased TiO$_2$ (DegussaTiO$_2$ P25) was used as reference material. According to Ohtani *et al*. [s1] Degussa P25 consists of 70% anatase and 30% rutile phase. Propyless (20% propylene glycol), which is a cutaneous emulsion, was purchased from Apoteket Kronan, Apoteket AB, Sweden.

**Synthesis 1 -Without acid catalyst**

4.8 g TTIP was mixed with 9.6 g 1-propanol and stirred for five minutes. 1.24 g water was mixed with 6.1 g 1-propanol and added to the alkoxide-solution at a rate of approximately 1 ml/min. During this time the solution went from colorless to white. The solution was split into two parts, one which was kept for analysis purposes and the other to which 0.5 g of MMC was added after approximately one hour of stirring to let the water react with the sol-gel reagents. The resulting mixture was stirred for 24 h under a plastic cover. The plastic cover was then removed and after one to two days a white precipitate was left in the bottom of the beaker.

**Synthesis 2 -With acid catalyst**

4.8 g TTIP was mixed with 9.6 g 1-propanol and stirred for five minutes. A mixture of 0.34 g HCl and 0.96 g 1-propanol was added to the alkoxide solution at a rate of approximately 1 ml/min. The resulting mixture was stirred for 30 min. A mixture of 1.24 g water and 6.1 g 1-propanol was then added to the alkoxide solution by approximately 1 ml/min. The solution was split into two equal parts where the first was kept for analysis purposes and the second one was mixed with 0.5 g of MMC powder. The resulting MMC- TiO$_2$ mixture was then stirred until gelation occurred after ~ 20 min. After gelation, the beaker was covered with a plastic cover for 24 h after which the cover was removed in order to let the gel dry. This resulted in a collapse to a white-transparent powder.

**Scanning electron microscopy (SEM)**

The SEM analyses were carried out on a Leo 1550 instrument (Zeiss). The sample powders were either pressed down in a piece of indium or put directly on carbon tape. The SEM-images were taken with the in-lens detector with 2 to10 kV acceleration voltage.

**Nitrogen gas sorption**

Nitrogen gas sorption isotherms were recorded with liquid nitrogen (77 K) using an ASAP 2020 (Micromeritics) instrument. Approximately 150 mg of sample was degassed at 100˚C under vacuum using a Micromeritic SmartVac Prep sample preparation unit for 10 h prior to analysis. The specific surface area (SSA) was determined by applying the Brunauer–Emmett–Teller (BET) method for adsorption values in the relative pressure range between 0.05–0.30 $P/P_0$ for the adsorption branch of the isotherm. The pore size distribution (PSD) was determined by the density functional theory (DFT) method using the model for nitrogen. The DFT analysis was carried out using the DFT Plus software from Micrometrics.

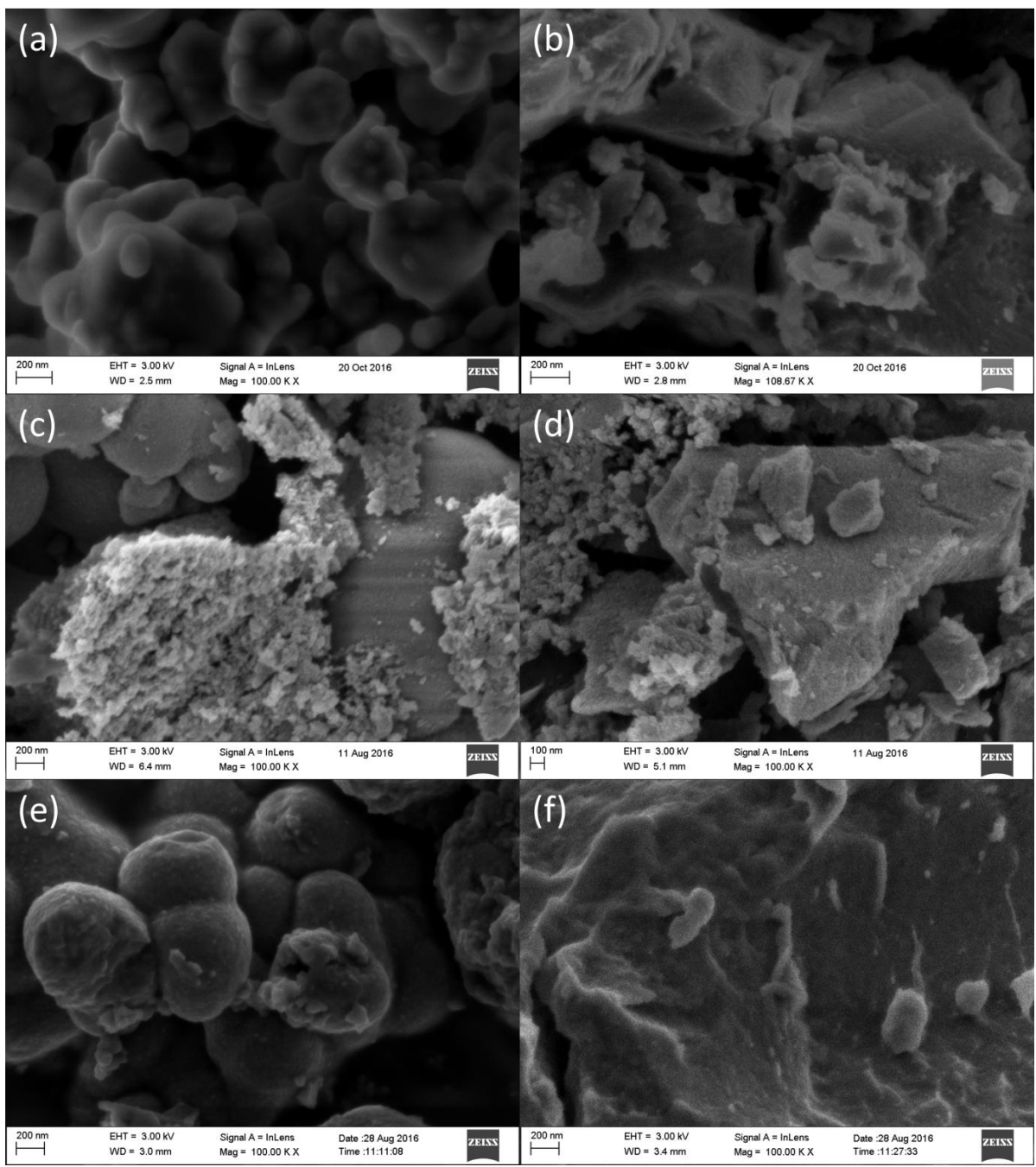

Figure S1. High resolution SEM images of a) Synthesis 1 and b) Synthesis 2 before heat treatment c) Synthesis 1 and d) Synthesis 2 after heat treatment e) Synthesis 1 and f) Synthesis 2 after 48 h of etching in 0.1 M $H_3PO_4$, in order to remove the MMC material.

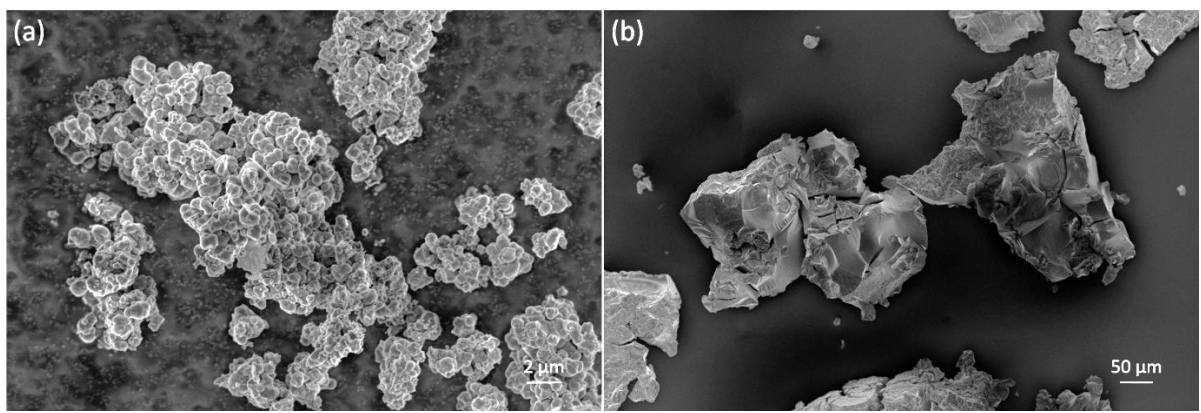

Figure S2. SEM images of lower resolution for a) Synthesis 1 and b) Synthesis 2 after heat treatment and 48 h of etching in 0.1 M $H_3PO_4$, in order to remove the MMC component.

Table S1. BET surface area and total pore volume of MMC (Upsalite), Synthesis 1 and 2 before and after heat treatment to 250 °C.

|  | Upsalite® | Synthesis 1 | Synthesis 1 after HT | Synthesis 2 | Synthesis 2 after HT |
|---|---|---|---|---|---|
| BET surface area [$m^2/g$] | 443 | 42 | 19 | 93 | 221 |
| Total pore volume [$cm^3/g$] | 0.99 | 0.14 | 0.10 | 0.10 | 0.26 |

Table S2. Active compounds in the commercial sunscreens A, B and C.

| Compounds | A | B | C |
|---|---|---|---|
| Titanium Dioxide |  | x | x |
| Butyl Methoxydibenzoylmethane | x | x | x |
| Drometrizole Trisiloxane |  |  | x |
| Ethylhexyl Triazone | x |  | x |
| Bis-Ethylhexyloxyphenol Methoxyphenyl Triazine |  | x | x |
| Octocrylene |  | x | x |
| Sodium Phenylbenzimidazole Sulfonate | x |  |  |